# Charge-Polarized Interfacial Superlattices in Marginally Twisted Hexagonal Boron Nitride

C. R .Woods[1,2]*, P. Ares[1,2], H. Nevison-Andrews[1,2], M. J. Holwill[1,2], R. Fabregas[1], F. Guinea[3,4], A. K. Geim[1,2], K. S. Novoselov[1,2,5,6], N. R. Walet[1], L. Fumagalli[1,2]*

*Email: colin.woods74@gmail.com, laura.fumagalli@manchester.ac.uk

[1]Department of Physics & Astronomy University of Manchester, Manchester, M13 9PL, UK
[2]National Graphene Institute, University of Manchester, Manchester, M13 9PL, UK
[3]Imdea Nanociencia, Faraday 9, Madrid 28049, Spain
[4]Donostia International Physics Center, Paseo Manuel de Lardizabal, 4, 20018 Donostia-San Sebastian, Spain
[5]Centre for Advanced 2D Materials, National University of Singapore, Singapore 117546, Singapore
[6]Chongqing 2D Materials Institute, Liangjiang New Area, Chongqing 400714, China

**When two-dimensional crystals are brought into close proximity, their interaction results in strong reconstruction of electronic spectrum and local crystal structure. Such reconstruction strongly depends on the twist angle between the two crystals and has received growing attention due to new interesting electronic and optical properties that arise in graphene and transitional metal dichalcogenides. Similarly, novel and potentially useful properties are expected to appear in insulating crystals. Here we study two insulating crystals of hexagonal boron nitride (hBN) stacked at a small twist angle. Using electrostatic force microscopy, we observe ferroelectric-like domains arranged in triangular superlattices with a large surface potential that is independent on the size and orientation of the domains as well as the thickness of the twisted hBN crystals. The observation is attributed to interfacial elastic deformations that result in domains with a large density of out-of-plane polarized dipoles formed by pairs of boron and nitrogen atoms belonging to the opposite interfacial surfaces. This effectively creates a bilayer-thick ferroelectric with oppositely polarized (BN and NB) dipoles in neighbouring domains, in agreement with our modelling. The demonstrated electrostatic domains and their superlattices offer many new possibilities in designing novel van der Waals heterostructures.**

One of the most promising avenues for controlling the properties of van der Waals (vdW) heterostructures is to adjust the angle between the stacked two-dimensional (2D) crystals. Such rotational control has allowed the observation of long-lived excitonic states[1], resonant tunnelling[2,3], and highly-correlated electronic states[4-7], including superconductivity in twisted bilayer graphene, among many others exciting effects. At the same time, the twist-depending properties of hexagonal boron nitride (hBN), one of the most used crystals for engineering vdW heterostructures, have been overlooked so far. Like in the case of graphene, atomically thin crystals of hBN can be obtained by exfoliating the bulk material down to monolayer thickness[8,9]. A wide-band insulator with strong polar covalent bonding between boron and nitrogen, hBN has proven itself indispensable for making high-quality vdW heterostructures[10,11] and lateral superlattices, especially in combination with graphene[9,12-17]. However, the possibility of using twisted hBN crystals has not been explored experimentally yet and, to the best of our knowledge, no spontaneous charge polarization has been observed within the increasingly rich variety of demonstrated vdW heterostructures.

In this work, we demonstrate that hBN crystals placed at an intentionally small ('marginal') twist angle create a superlattice of charge-polarized macroscopic domains confined to the interface. The system undergoes reconstruction into a periodic commensurate phase that results in a high density of polarized interfacial dipoles between the hBN layers at the interface, as measured by electrostatic scanning probe microscopy[18] at room temperature. We show that two dominant crystal alignments at 0° and 180° angle, referred to as parallel and antiparallel respectively, experience different reconstruction. Only the parallel alignment gives rise to ferroelectric-like domains, where aligned boron and nitrogen atoms at the interface layers create a dipolar field that reverses its sign in adjacent domains. Our conclusions are strongly supported by calculations of atomic reconstruction and charge density in the interfacial layer.

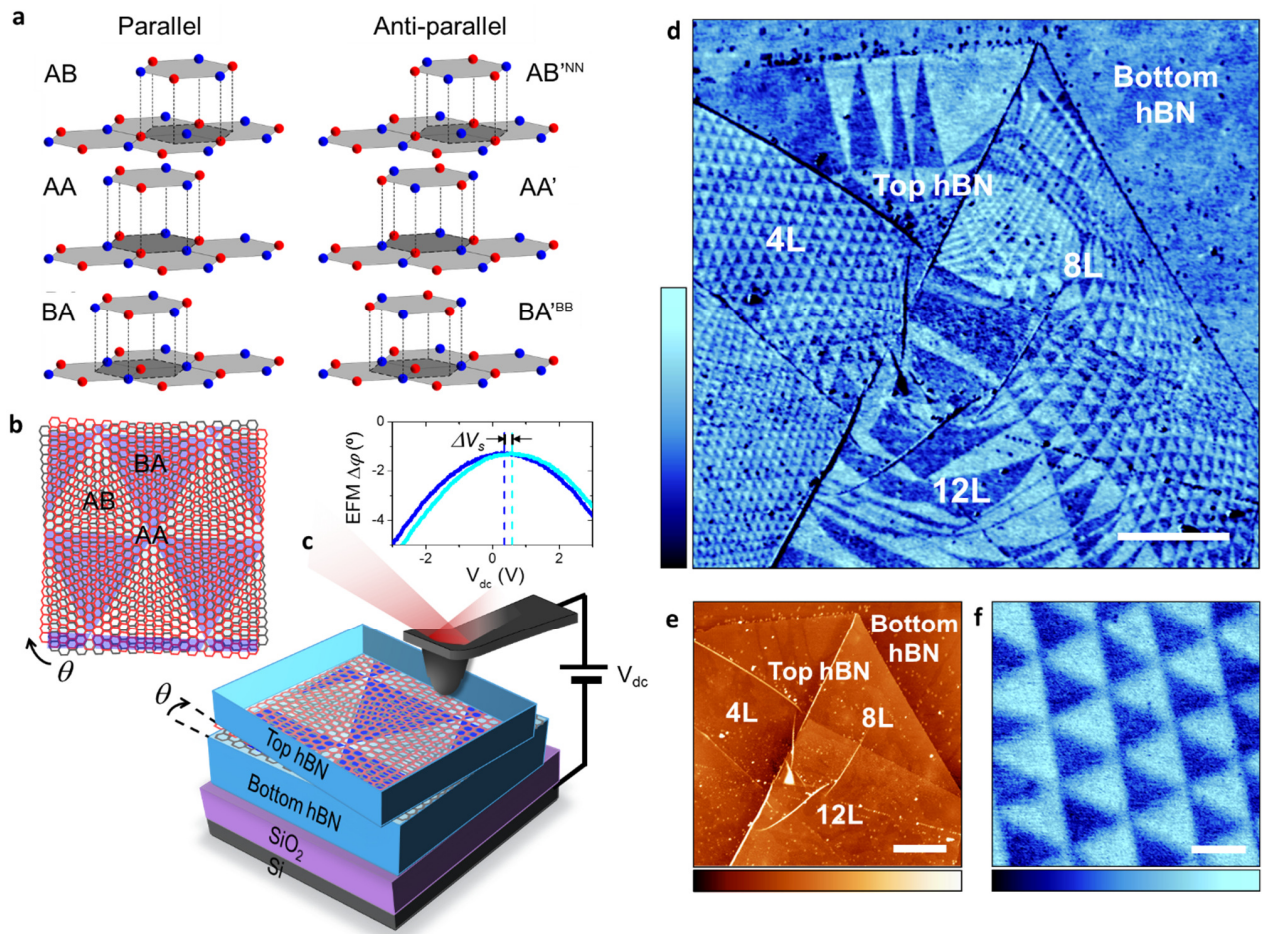

**Figure 1 | Electrostatic imaging of charge polarization in marginally twisted hBN. (a)** Illustration of six high-symmetry stacking configurations for the hBN-hBN interface. Nitrogen atoms are shown in red; boron atoms in blue. **(b)** Schematic of adjacent hBN atomic layers (red and grey) misaligned by a small angle, $\theta$. Dark and light triangles represent predominantly AB and BA regions. **(c)** Schematic of our experimental setup. Red and grey hexagonal lattices are in the top and bottom hBN, respectively. A voltage bias is applied between the AFM probe and the silicon substrate. Inset: representative dc-EFM curves as a function of the applied dc bias in two adjacent triangular domains. The horizontal shift of the maximum of the curves yields the variation in surface potential, $\Delta V_s$, between the domains. **(d)** Representative dc-EFM image of twisted hBN showing large areas with triangular potential modulation. Changes in domains' shape and periodicity are due to small changes in $\theta$ caused by irregular strain and the wrinkles seen in the corresponding AFM topography image in **(e)**. The top hBN crystal has 4-, 8- and 12-layer thick regions. **(f)** Zoom-in of a region in **(d)** with regular domains. Scale bars: **(d,e)** 2 μm; **(f)** 200 nm. Colour bars: **(d,f)** 10° ; **(e)** 12 nm.

Our heterostructures were made of a thin hBN crystal (1 - 20 layers) placed on top of a generally thicker hBN substrate (> 30 layers) on highly doped Si/SiO₂ substrates, following the procedures detailed in Methods (also, see Supplementary Figs. S1 and S2). The two crystals were aligned at a twist angle $\theta < 1°$ and, for convenience, are referred to as top and bottom hBN, as depicted in Fig. 1b. We characterized the heterostructures using atomic force microscopy (AFM) and its electrostatic modes, electrostatic force microscopy (EFM)[19] and Kelvin probe force microscopy (KPFM)[20], which allow visualization of local electric charges by detecting variations in the surface potential (Methods and SI). EFM can be implemented with either dc or ac voltage bias between the tip and the Si substrate, which we refer to as dc-EFM[21] and ac-EFM[22], respectively (Supplementary Fig. S3). Below we focus on results obtained using dc-EFM for its high spatial resolution and simplicity of implementation. In this case, electrostatic domains can be detected directly in the AFM phase image by applying a dc bias, which allows faster scanning as compared to other electrostatic modes. ac-EFM and KPFM, which are generally slower and more demanding (they require to record an additional image using a second lock-in and, in the case of KPFM, a second feedback control[20]), yielded the same information (see Supplementary Figs. S4-S8).

Figure 1 shows representative images taken from one of our twisted-hBN samples, in which the top hBN crystal has regions of 4-, 8- and 12-layer thickness (also see Supplementary Fig. S2). No superlattice pattern could be detected in the topography image (Fig. 1e) using standard contact and intermittent-contact modes. On the other hand, the corresponding dc-EFM image (Fig. 1d) reveals a clear pattern with triangular domains that are periodically arranged in many regions. The observed patterns do not extend beyond the area covered by the top hBN, indicating that they originate at the interface between the hBN crystals. This is also consistent with the fact that no pattern was detected on bubbles with contamination trapped between the two hBN crystals[23] (Supplementary Figs. S4 and S5). Note that EFM-imaging is sensitive to subsurface properties[24,25] (in our case, this is the interface between the hBN crystals) thanks to the long-range nature of electrostatic forces. The marginally twisted hBN shows large areas with triangular domains (Fig. 1f). However, as the size of domains increases, the moiré pattern becomes irregular. This can be understood by noticing that wrinkles seen in our AFM images are likely to induce local inhomogeneous strain and, for small twist angle ($\theta < 0.5°$), even minor strain variations lead to large changes ($\propto 1/\theta$) in the moiré periodicity[26]. Upon changing the sign of the applied dc bias, the contrast inverted (Supplementary Fig. S7), indicating that the observed pattern originates in permanent, built-in surface charges (as opposed to changes in dielectric properties[22,27], see Supplementary Fig. S5). This was further confirmed by taking dc-EFM curves as a function of the bias (Fig. 1c, inset). The curves measured on adjacent domains are shifted with respect to each other by the potential difference, $\Delta V_s$ (see Methods and SI), which was found to be the same $\Delta V_s \approx 240 \pm 30$ mV in all our samples, irrespectively of the domain size (down to ~ 30 nm), shape and orientation (Fig. 3f). Additional KPFM images confirmed these results (see Supplementary Fig. S8).

The triangular patterns observed in our twisted-hBN structure resemble those observed in twisted bilayer graphene[28-30]. In the latter case, the triangular domains are alternating regions of AB and BA stacking. The case of hBN is more complex because its unit cell has two nonidentical sublattices, which leads to two cases of perfect alignment, parallel and antiparallel. Figure 1a illustrates the resulting six high-symmetry stacking configurations for an hBN-hBN interface. The parallel configuration allows AA, AB, and BA domains. AB and BA are equivalent, except for the layer inversion such that the boron atom align with the nitrogen atom residing in either top or bottom layer, respectively (Fig. 1a). Because AB/BA alignment is more energetically favourable than AA[31,32], it is expected that, if the local crystal strain is allowed to relax, triangular domains in Fig. 1b should have predominantly AB and BA stacking. The antiparallel configuration can form AA', AB'$^{NN}$ and BA'$^{BB}$ domains, where AB'$^{NN}$ and BA'$^{BB}$ refer to the stacking where interfacial nitrogen is aligned with nitrogen and boron with boron, respectively (Fig. 1a). For small twist angles, regions of mixed stacking may occur but, as AA' stacking is more favourable than AB'$^{NN}$ and BA'$^{BB}$,[31,32] once the system relaxes, we expect hexagonal domains of predominantly AA' stacking (see below).

To determine whether the observed domains are a result of the parallel or antiparallel interfacial configuration, we fabricated and studied marginally-twisted heterostructures with monolayer steps on the surface of the bottom hBN crystal. Crystallographically, this guarantees the presence of both parallel and antiparallel alignments within a single heterostructure device (Fig. 2a). Because bulk hBN naturally has AA' stacking, a monolayer terrace effectively produces rotation by 180° with respect to the adjacent region, as illustrated in Fig. 2a. Therefore, by aligning the top hBN over the single-layer terrace, one can probe both parallel and antiparallel configurations at the same sample, as illustrated in Fig. 2a - BA stacking on the left side and AB'$^{NN}$ stacking on the right side. Figures 2b and 2c show AFM and EFM images for a sample with top-hBN aligned over a monolayer step indicated with the dashed yellow line. The topography image in Fig. 2b shows that the terrace is formed by a monolayer ($h \approx 0.33$ nm). In the corresponding dc-EFM image (Fig. 2c), we observe a clear triangular pattern on one side of the terrace which cuts off sharply at the terrace edge, with no periodic signal on the other side. This suggests that periodic domains are possible only for one orientation type, either parallel or antiparallel. We repeated this experiment on several samples, confirming the generality of the observation (Supplementary Figs. S4 and S6). To corroborate this further and rule out the possibility that monolayer terraces facilitate a rotation or some other changes that prevent the observation of the charge-polarized domains on one of the sides of monolayer steps, we also fabricated samples with bilayer terraces ($h \approx 0.66$ nm). This crystallographic arrangement is shown schematically in Fig. 2d. It is clear that bilayer steps should make alignment identical in the neighbouring regions (Fig. 2d shows the case of BA stacking). In agreement with the expectations, we found the ferroelectric domains on both sides of bilayer steps with no difference in the contrast (Figs. 2e and f).

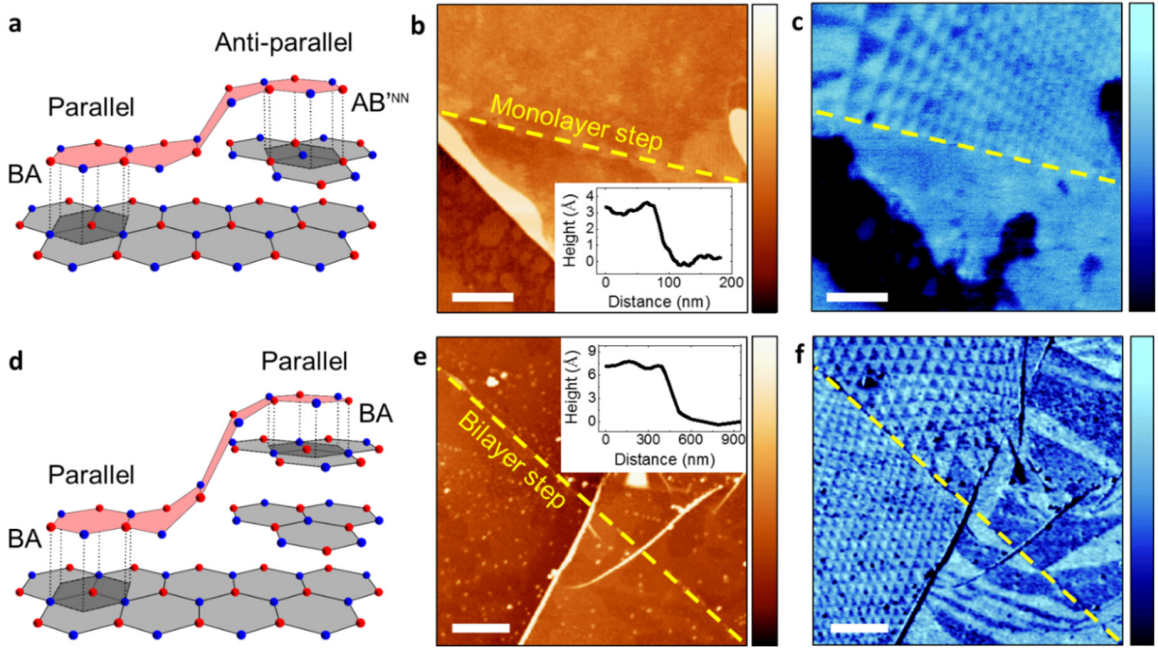

**Figure 2 | Effect of mono- and bi- layer terraces on occurrence of charged-polarized domains. (a)** Illustration of hBN alignment over a monolayer terrace in the bottom hBN. The terrace forces an alignment change from parallel (left) to antiparallel (right) at the interface between top hBN (light red) and bottom hBN (light grey; AA' stacking). Dark-grey areas indicate BA and AB'$^{NN}$ stacking. **(b)** AFM topography image of a representative sample, showing an hBN bilayer crystal covering a monolayer terrace in the bottom hBN. Inset: height profile across the step. **(c)** Corresponding dc-EFM image. The triangular potential modulation is visible only on one side of the step, marked by the yellow dashed lines in **(b)** and **(c)**. **(d)** Schematic as in **(a)** but for a bilayer terrace. The terrace in the bottom hBN (AA' stacking) does not influence the parallel alignment of the top hBN. The dark-grey shaded areas indicate BA stacking. **(e)** AFM topography image of an hBN crystal covering a bilayer step in the bottom crystal (inset: the step profile). **(f)** Corresponding dc-EFM image. The triangular modulation is visible on both sides of the step marked in yellow. Scale bars: **(b,c)** 250 nm; **(e,f)** 500 nm. Colour bars: **(b,e)** 12 nm; **(c,f)** 10°.

It is important to note that the observed potential contrast is independent of the superlattice periodicity, as shown in Fig. 3f. The contrast remained constant down to the smallest domains, of approximately 30 nm in size, that we could detect within our lateral resolution (Methods). This observation rules out the possibility that the potential pattern can arise from piezoelectricity effects as a result of in-plane deformation, as reported for hBN previously[9]. Indeed, the piezoelectric charge should be proportional to the strain gradient[33] and, accordingly, strain variations are expected to decrease with increasing the domain size, in contrast to the experimental results. Our analysis also shows that strain gradients and thus the piezoelectric charge should be localized at the domain edges (Supplementary Fig. S9). Furthermore, the observed contrast was found to be practically independent of the thickness of top and bottom hBN crystals for the investigated range of thicknesses. We detected the contrast in all our samples with top hBN ranging from a monolayer up to 18 layers (6 nm) whereas bottom hBN was up to 80 nm thick. This independence is of practical importance because it greatly simplifies access to the charge-polarized twisted superlattices, without the need of using mono- or few- layer crystals, which is experimentally challenging for the case of hBN[8]. We note, however, that for thick hBN crystals, atomic relaxation at the interface is generally expected to be hindered by an additional elastic contribution from the bulk, which may result in weaker interfacial strain and somewhat smaller polarization (see below).

To quantify our experimental observations, we calculated superlattice reconstructions at the twisted hBN interface, focusing on the case of marginally twisted monolayers. Atomic reconstructions occur as a result of the balance between the energetically favourable interlayer alignment and the elastic energy required to reach the relaxed lattice configuration. Figures 3a and 3b show the stacking order after relaxation for a twist angle of 0.33° for parallel and antiparallel alignments, respectively. In all cases the observed superlattices form periodic commensurate states with superlattice periodicity matching that of the moiré pattern between them. The antiparallel alignment forces the superlattice to relax into a rather complex configuration with threefold symmetry (Fig. 3b), maximising the area of AA' stacking between the two crystals whilst minimising the unfavourable AB'$^{NN}$ and BA'$^{BB}$ stacking. In contrast, the parallel interfacial alignment relaxes into a much simpler triangular superlattice (Fig. 3a), where the AA aligned region becomes very small. The latter maximises the total area of the inversion-symmetric AB/BA stacking that is energetically favourable with respect to AA stacking[32]. Comparison between the calculated shapes and our EFM images strongly suggest that the observed superlattices represent the case of parallel alignment with alternating AB and BA domains. Next, we calculated a charge density distribution within the domains. This required full details of all energy bands in our tight binding model (details in SI). The calculations show that even without lattice reconstruction a considerable charge polarization occurs at the interface (Fig. 3c). However, to obtain triangular domains with sharp boundaries such as those observed in the experiment, the lattice relaxation is needed to be accounted for, as it enhances crystallographic alignment over the entire unit cell (Fig. 3d). A large charge-density modulation was found only for the parallel alignment (Figs. 3 d,e). For the antiparallel one, our calculations yielded extremely small surface charge densities, three orders of magnitude smaller than in the case of parallel alignment. This small polarization would be dwarfed by piezoelectric effects induced by in-plane elastic strain. The strong potential modulation observed experimentally is another proof of the parallel rather than antiparallel alignment in marginally twisted hBN.

The interaction-induced charge density within hBN monolayers is not expected to vary with the domain size, in agreement with the experiment (as long as it is larger than the reconstructed region at the domain wall, approximately 4 nm according to our calculations). On the other hand, the absolute value of the surface charge was found to be very sensitive to interlayer hopping parameters. For the range of values reported in literature, we obtained charge densities around few $10^{12}$ cm$^{-2}$ (see Figs. 3d,e). For the larger hopping parameters used in Fig. 3e, the density modulation reaches approximately ±3 mCm$^{-2}$. Using the simple parallel-plate approximation and assuming interlayer dielectric constant of 1, the above value translates into a surface potential modulation of about 0.2 eV, in fair agreement with the experiment. Nonetheless, the comparison between the experiment and theory in Fig. 3f indicates that the theory probably underestimates the strength of interlayer interactions. Further work using advanced computational methods is required to clarify the interaction strength and the induced spontaneous charge polarization at the interface.

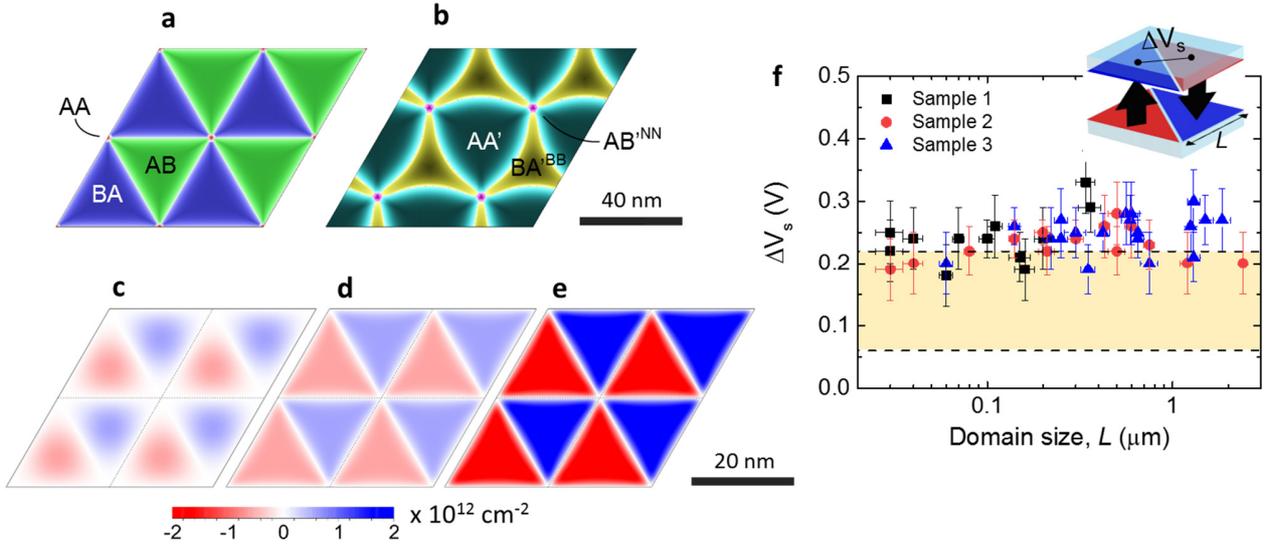

**Figure 3 | Calculated charge-density distribution in marginally twisted hBN. (a,b)** Dominant stacking order for twisted-bilayer hBN, calculated as in Refs. [34,35], for $\theta = 0.33°$ in the case of parallel **(a)** and antiparallel **(b)** alignment. AB staking is shown in dark green, BA – dark blue, AA – red, AA' – dark cyan, BA'$^{BB}$ – dark yellow, and AB'$^{NN}$ – magenta. The AA and BA'$^{BB}$ alignments occur at the intersections of the AB and BA regions, and AA' and BA'$^{NN}$ regions, respectively. The colour intensity indicates the degree of alignment of boron and nitrogen atoms located in the two hBN monolayers. The atoms are perfectly aligned in the domains' centres. Scale bar: 40 nm. **(c,d,e)** Charge density distribution within individual hBN monolayers, which is induced by interlayer interaction for the case of parallel alignment for $\theta = 0.52°$. Scale bar: 20 nm. **(c)** Relatively weak interlayer hopping without lattice relaxation. **(d)** Same hoping but accounting for the lattice relaxation. **(e)** Stronger hopping with lattice relaxation. Twisted bilayer hBN remains charge-neutral, and the charge polarity is reversed between the two layers (red and blue reverse), which also reflects the inversion symmetry of AB/BA stacking. **(f)** Electrostatic potential variation in the centre of AB and BA, as illustrated in the inset. The experimental values (symbols) as a function of domain size. Within our accuracy, the potential is size-independent. The yellow-shaded region denotes the calculated surface potential values delimited by the two hopping amplitudes used in **(d,e).**

To qualitatively understand the physics behind our observations, let us consider aligned pairs of boron and nitrogen atoms located in the different hBN monolayers of the interface. Their interaction generates BN and NB dipoles aligned along hBN c-axis (out of plane). For AA' stacking (Fig. 1a), the dipoles have both polarizations (BN and NB) within the same unit cell. In contrast, for AB/BA stacking, only one orientation of dipoles (either BN or NB) is present, which breaks the symmetry along the c-axis and effectively creates a ferroelectric bilayer with a fixed polarization. The reversal between AB and BA stacking in adjacent domains leads to their opposite charge polarization, which is responsible for the surface potential difference imaged by EFM and KPFM.

In conclusions, we observed triangular dipolar domains in marginally twisted hBN that originate at the interface between the two hBN crystals, as in a bilayer-thick ferroelectric in the out-of-plane direction. The demonstrated charge-polarized superlattices provide a fascinating platform to study superlattice phenomena and enable new architectures for vdW heterostructures. The fact that a small twist angle between two insulating 2D crystals can generate a strong interfacial charge polarization of known amplitude and periodicity is a new important addition to the arsenal of vdW technologies. It can be used, for example, to create an artificial surface potential and modify properties of adjacent 2D materials such as graphene or transition metal dichalcogenides, and to engineer novel electronic and optical devices (ferroelectric memories, light-emitting diodes, interfacial 2D electron gases, etc).

**Methods**

**Sample preparation**. We fabricated twisted-hBN heterostructures on oxidized Si wafers (290 nm of $SiO_2$) by using the PDMS/PMMA (Polydimethylsiloxane/Poly(methyl methacrylate) dry transfer technique[36], described briefly here and further in SI. In short, we exfoliated hBN crystals onto the Si substrate and identified target crystals using optical microscopy (Supplementary Fig. S2). We chose the crystals on the basis of two requirements: first, two hBN crystals should be adjacent to each other and, second, the perspective bottom hBN should have a monolayer or bilayer terrace in its top surface. The first condition assured near-perfect alignment between the two crystals using their zigzag to zigzag or armchair to armchair edges. Indeed, if the crystals are adjacent to each other, they are likely to split from the same original crystal. Without following this rule, there would be a 50% chance of creating 30° misaligned samples (zigzag to armchair alignment). The second requirement was used in the experiments discussed in Fig. 2, where we studied parallel and antiparallel alignment in a single device.

**AFM and electrostatic imaging.** We acquired simultaneous AFM topography and electrostatic images using the standard two-pass method[18]. First, we measured topography in the dynamic mode by oscillating the tip at its free mechanical resonance with no applied electric field. In the second pass, the electrostatic signal was recorded while retracing the topography line with an applied voltage after retracting the tip a few nm. We measured the electrostatic signal with either dc or ac applied voltage. In dc-EFM[21] (Supplementary Fig. S3a), also called phase-EFM, we applied a dc bias of 2-3 V and recorded the AFM phase shift, $\Delta\varphi$, of the cantilever mechanical oscillations, which depends quadratically on the surface potential, $V_s$. In ac-EFM[22] (Supplementary Fig. S3b), we excited the cantilever with an ac voltage of amplitude 4-5 V and frequency $\omega$ = 1-10 kHz, and we recorded the amplitude variation of $\Delta\varphi$ at $\omega$ and $2\omega$ using an additional lock-in amplifier. No triangular domains were detected at $2\omega$, indicating that the observed domains do not have a dielectric origin[22,27] (Supplementary Fig. S5). We quantified the surface potential variation, $\Delta V_s$, in Fig. 3f by taking $\Delta\varphi$ versus dc voltage curves in the centre of the domains, as detailed in SI. KPFM images (Supplementary Fig. S8) were recorded by nullifying the amplitude of $\Delta\varphi$ at $\omega$ using a second feedback control and applying both dc and ac voltages, where the dc voltage is used to compensate a variation in $V_s$ and thus obtain $\Delta V_s$ images. All the curves and KPFM images were taken with the tip at minimal scan height, resulting to be in the range 8-12 nm from the dipolar plane. This is important, because the built-in potential detected by the tip reduces with the tip-surface distance due to long-range forces from the tip cone and cantilever. We note that the force-gradient measurement approach employed here based on phase-shift detection is advantageous to minimize the impact of such forces as compared to conventional amplitude-modulation EFM or KPFM that measure the force. In addition, we employed doped silicon probes (Nanosensors PPP-CONTR, spring constant 0.5 – 1.5 $Nm^{-1}$) that have smaller tip radii (< 7 nm) than conventional metal-coated probes. We thus estimate the lateral resolution in our experiments to be in the range 5-10 nm[37]. This agrees with our observations in which we quantified the built-it potential to be constant for triangular domains as small as 30 nm. Care should be taken if electrostatic images are taken with larger tip radii and scan heights or on thicker top-hBN, as the observed built-in potential may be reduced due to tip-sample force convolution. Data were acquired and processed using WSxM software[38].

**Theoretical calculations.** The theoretical calculations were performed in two stages: first, we used LAMMPS[39] to minimize the energy using a classical potential model for relaxation[34], using the 'inter-layer potential'(ILP) from Refs.[40,41] with the Tersoff in-layer potential[42,43]. The results were plotted using an extension of the method in Ref. [34] where we took into account all six alignment options shown in Fig. 1. Using the found deformed lattice, we then performed tight-binding calculations using electronic coupling with exponential Koster-Slater interlayer hoppings. We used the standard in-layer nearest neighbour hopping of 2.33 eV[32], neglecting its modification by minor bond stretching. The charge density was then calculated by summing over all occupied states, which limited the smallest angle we could perform calculations for to about 1°. Details of these calculations as well as further results on the electronic structure of twisted hBN can be found in Ref.[35].

**Acknowledgements**

We acknowledge support from EU Flagship Programs (Graphene CNECTICT-604391 and 2D-SIPC Quantum Technology), European Research Council Synergy Grant Hetero2D, the Royal Society, EPSRC grants EP/N010345/1, EP/P026850/1, EP/S030719/1. P. A., H. N-A., R. F. and L. F. received funding from the Marie Sklodowska-Curie Actions (grants 793394, 842402) and the European Research Council (grant 819417) under the European Union Horizon 2020 research and innovation programme.

# Supplementary Information

**S1 Sample fabrication**

Twisted-hBN heterostructures were fabricated as schematically shown in Fig. S1. First, the hBN crystals were isolated by mechanically exfoliating commercially available bulk crystals. The mechanical exfoliation was done using scotch tape to reduce the bulk material until the crystals were significantly thinner and more numerous. The resultant material was then deposited onto a $SiO_2$/Si (290 nm) substrate. $SiO_2$/Si substrates provide excellent optical contrast for even the thinnest flakes and terraces on their surfaces, particularly when combined with wavelength filtering, dark-field imaging and Nomarski filtering (see Fig. S2). These techniques were used to identify pairs of adjacent hBN crystals: one thin crystal of 1-20 layers (top hBN) and one thicker crystal of more than 30 layers (bottom hBN). Importantly, the bottom hBN crystal was only used if it contained a monolayer or bilayer terrace in its surface (Fig. S2b), as confirmed by AFM topography imaging (Fig. S2c and Figs. 2b and 2e in the main text). Further, using adjacent hBN crystals removes the requirement to use crystallographic fractures as an alignment tool during the transfer process (a method limited to 50% probability of success). Because it is likely that the two crystals are from the same growth domain, they are already in near-perfect alignment. Once the pair of hBN crystals were found on the $SiO_2$/Si substrate , we brought them together using the PDMS/PMMA (Polydimethylsiloxane/Poly(methyl methacrylate)) dry-peel transfer technique[36]. The bottom hBN (light grey) and top hBN (light red) are identified on the $SiO_2$/Si substrate and a PDMS/PMMA membrane (green) is positioned above the relevant area using a micromanipulation stage (Fig. S1a). The membrane is brought into contact with top hBN only (Fig. S1b). Then the membrane is removed, lifting the top hBN crystal with it (Fig. S1c). The membrane was then translated with no rotation, so that the top hBN crystal was above the terrace in the bottom hBN, and the two were brought into contact (Fig. S1d). Finally, the membrane was removed, leaving the top hBN on the bottom hBN (Fig. S1e).

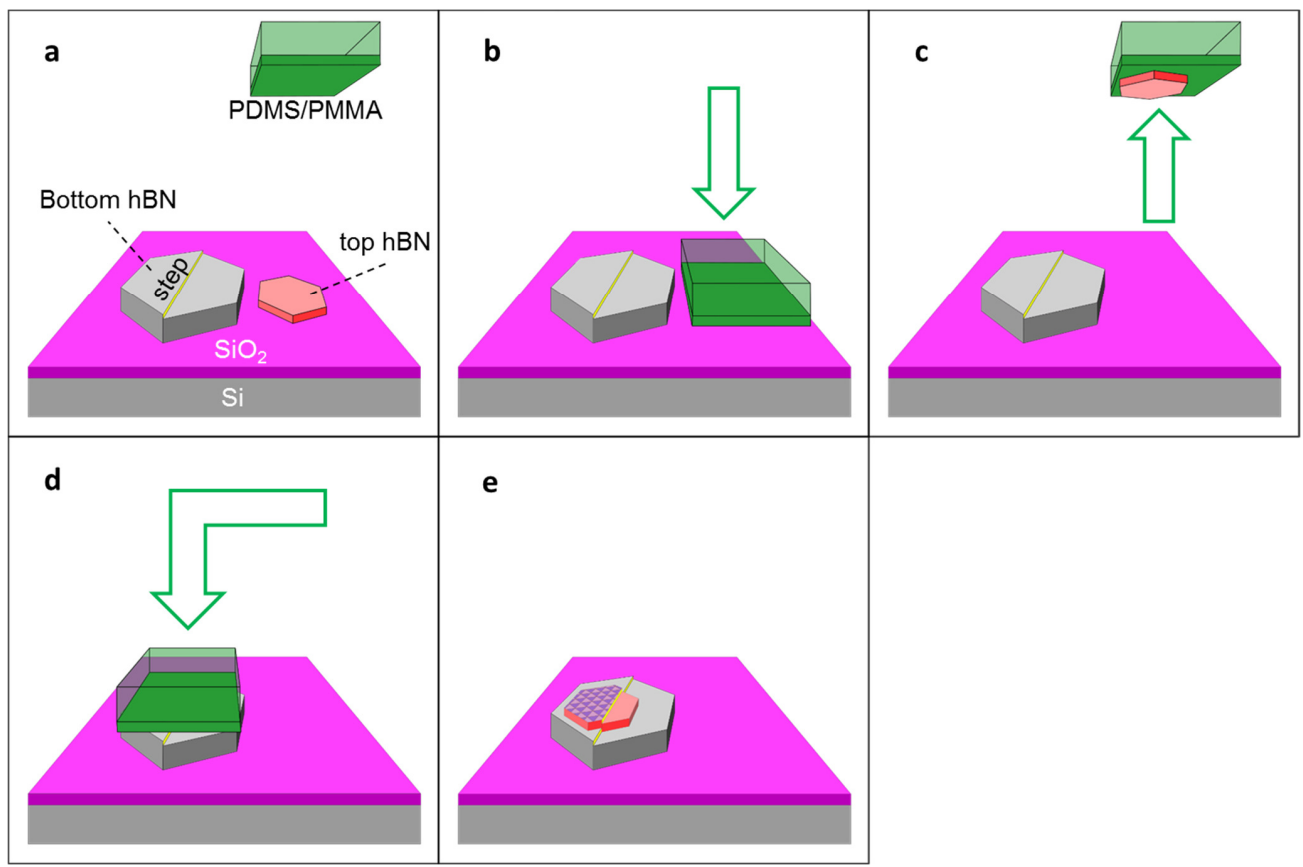

**Figure S1 | Schematic of the procedure used to fabricate marginally twisted hBN heterostructures. (a)** Bottom (light grey) and top (light red) hBN crystals are identified on a SiO$_2$/Si substrate. The bottom hBN is confirmed to have a monolayer or bilayer terrace in its surface (yellow step). **(b)** A PDMS (faint green) and PMMA (green) membrane is brought into contact with the top hBN. **(c)** The membrane and top hBN are lifted from the substrate. **(d)** The membrane is translated above the bottom hBN with no rotation and brought into contact. **(e)** The membrane is removed leaving both crystals in place.

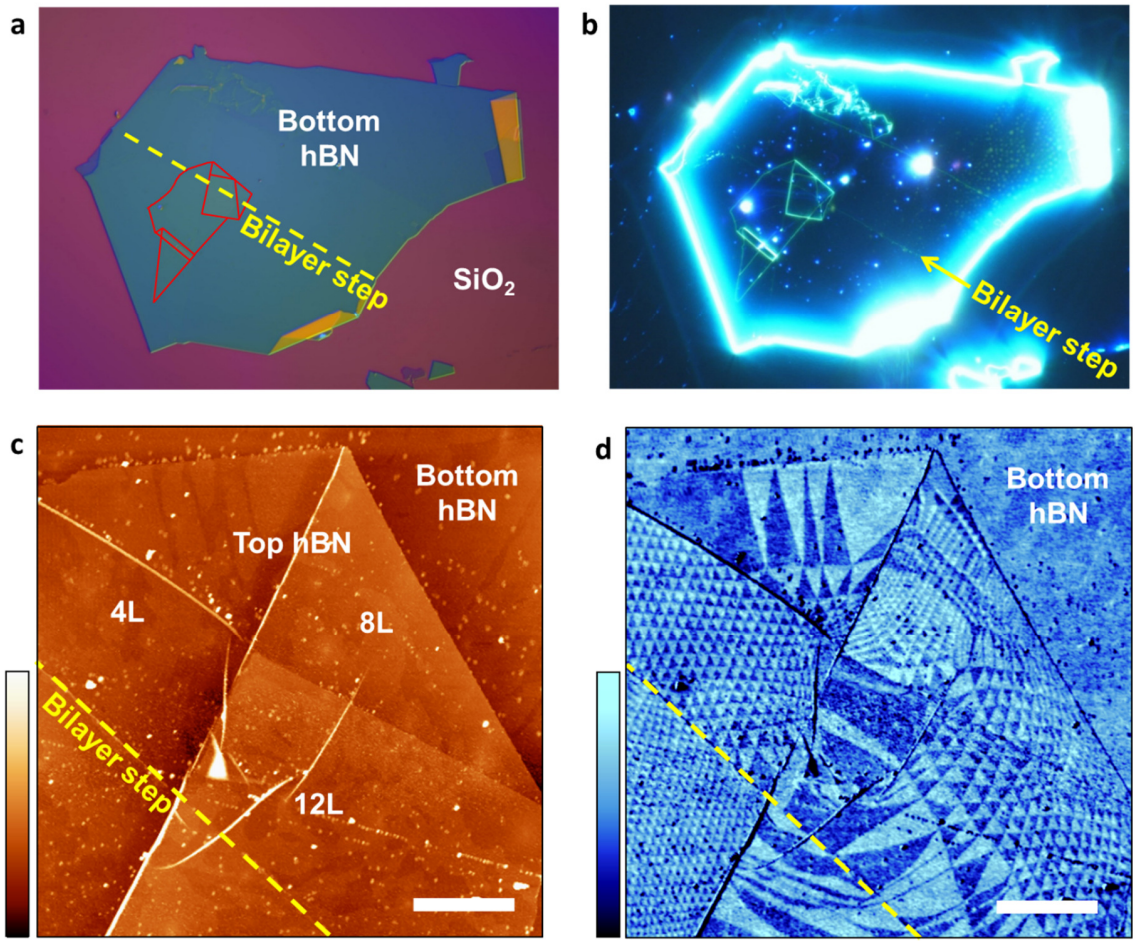

**Figure S2 | Optical and AFM images of marginally twisted hBN on terraces. (a,b)** Representative optical images of one of our twisted samples, shown in Fig. 1d-f and Fig. 2e,f in the main text. The top hBN crystal (red line) was transferred on the bottom hBN crystal in a region with a bilayer step, marked by the yellow dashed line, which is visible in the dark-field image in **(b)**. **(c,d)** Corresponding AFM topography and EFM image of the top hBN in the region near the bilayer step (yellow dashed line). The top hBN crystal has 4-layer, 8-layer and 12-layer-thick regions. The triangular potential modulation is detected in all the regions, regardless of their thickness, and on both sides of the bilayer step. Scalebar: **(c,d)** 2 μm. Colour bar: **(c)** 12 nm; **(d)** 10°.

## S2 Electrostatic imaging

The triangular potential modulation was detected using electrostatic force microscopy (EFM) and Kelvin probe force microscopy (KPFM). Both techniques are non-contact scanning probe techniques that probe the electrostatic interaction between a conductive AFM tip and the sample[18,20]. EFM measures local electrostatic force variations that can originate from either a variation in the surface potential or in the dielectric properties of the sample. In this section, we present additional information and images using EFM. KPFM is just an advanced EFM mode in which surface potential variations are recorded using an additional feedback loop, and we discuss it in the next section (see S3).

EFM images in the main text were taken using dc-EFM mode (also known as phase-EFM)[21] by simply applying a dc voltage, as illustrated in Fig. S3a. The cantilever was oscillated at its free mechanical resonance with a dc voltage applied between the tip and the silicon substrate, and the phase shift, $\Delta\varphi$, of the mechanical oscillation of the cantilever was recorded while scanning the surface. The electrostatic force experienced by the cantilever can be written as $F_{el}(z) = \partial C/\partial z\, (V_{dc} - V_s)^2/2$, where $C$ is the total tip-sample capacitance, $V_{dc}$ is the applied dc voltage between the AFM tip and the sample substrate, $V_s$ is the surface potential and $z$ is the tip-surface distance. To a first approximation, the phase shift directly depends on the force gradient as $\Delta\varphi(z) = -\frac{Q}{k}\frac{\partial F}{\partial z}$. Thus, in the presence of an electrostatic force, it can be written as $\Delta\varphi(z) = -\frac{Q}{2k}\frac{\partial^2 C}{\partial z^2}(V_{dc} - V_s)^2$ where $Q$ is the quality factor of the cantilever and $k$ its spring constant. Hence, the phase shift varies proportionally to the square of the tip-surface potential difference and to the second derivative of the tip-surface capacitance, $\partial^2 C/\partial z^2$. The latter is a complex function of the geometric and dielectric properties of the tip-sample system.

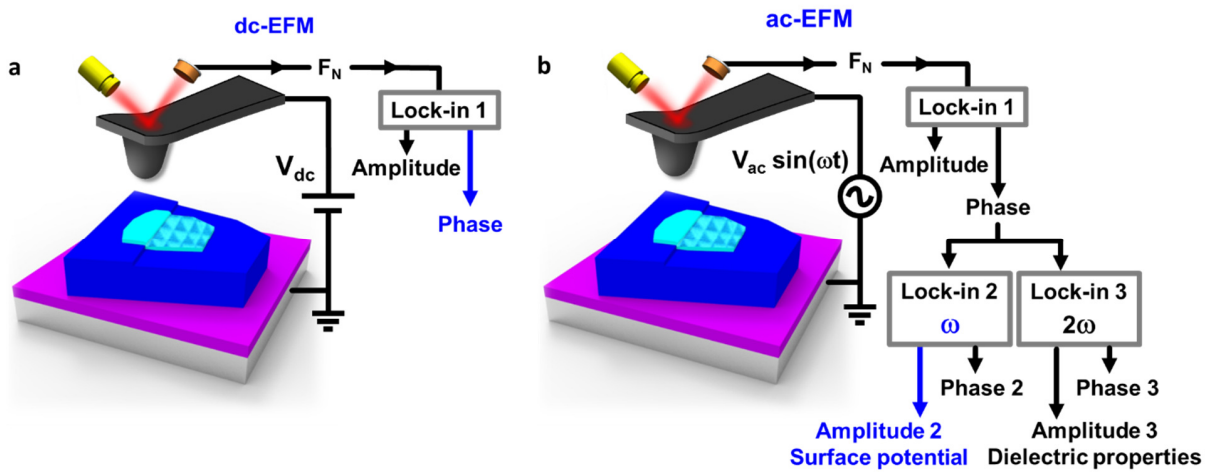

**Figure S3 | Schematics of the EFM setup**. **(a)** In dc-EFM, the AFM phase was recorded during the second pass with a dc voltage applied between the AFM probe and the silicon substrate. **(b)** In ac-EFM, an ac voltage of frequency $\omega$ was applied between the AFM probe and the silicon substrate during the second pass. The amplitude of $\Delta\varphi(\omega)$, which depends on the surface potential, showed periodic triangular modulation on the twisted-hBN crystals, while the amplitude of $\Delta\varphi(2\omega)$, which depends only on the surface dielectric properties, showed no periodic pattern.

We recorded $\Delta\varphi$ using the standard two-pass method. First, we acquired the topography image with no applied voltage. Then we retraced it with an applied dc bias of 2-3 V, the AFM feedback control switched off and the tip lifted up a few nm with respect to the first pass. The lift height, $z_{lift}$, set in the range 3-5 nm, was chosen large enough to avoid short-range forces, but as small as possible to minimize the tip-surface distance, $z$. This is important to maximize the lateral resolution of the technique, which is set by the $\partial^2 C/\partial z^2$ term and decreases with the tip moving away from the surface. By keeping the oscillation amplitude in the range of 5-10 nm, we typically took the EFM images at a total scan height of 8-20 nm from the interface between the twisted hBN crystals. We note that by probing the force gradient through $\Delta\varphi$ instead of the force as in standard amplitude-modulation EFM or KPFM, the dc-EFM mode employed here is less sensitive to long-range forces from the tip cone and cantilever. Thus, it allows higher lateral resolution and it is advantageous here to study small domains. To discriminate whether the observed triangular pattern was a built-in potential or a change in the dielectric properties of the heterostructure, we took $\Delta\varphi$ curves as a function of the applied dc bias in the centre of the triangular domains, as detailed in section S3. They show the expected parabolic behaviour (see Fig. S7)[44-46]. While the curvature, set by the capacitive term $\partial^2 C/\partial z^2$, was independent of the domain, the maximum of the parabola shifted a few hundreds of mV with the domain polarity. This allowed us to conclude that the triangular modulation detected in the EFM images is a built-in potential due to the interfacial charge distribution between twisted hBN crystals.

To support this conclusion, we also took images in ac-EFM mode, illustrated in Fig. S3b, which allowed us to record dielectric images. In this mode, an ac voltage bias of frequency $\omega$ and amplitude $V_{ac}$ was applied between the AFM tip and the sample substrate. This modulates the electrostatic force at $\omega$ and $2\omega$ with amplitude $F_{el}(\omega) = \partial C/\partial z \, V_s \cdot V_{ac}$ and $F_{el}(2\omega) = \partial C/\partial z \cdot V_{ac}^2/4$, respectively. While the $\omega$ harmonic is again proportional to both the surface potential $V_s$ and the dielectric properties through the capacitive term $\partial C/\partial z$, the $2\omega$ harmonic depends only on the $\partial C/\partial z$ term[27]. Using two additional lock-in amplifiers, we measured the amplitude of both harmonics. Note that, also in this mode, we measured the phase shift $\Delta\varphi$ of the mechanical oscillation of the cantilever instead of the force, thus recording the two phase harmonics, $\Delta\varphi(\omega)$ and $\Delta\varphi(2\omega)$. Again, this is advantageous because $\Delta\varphi$ is proportional to $\partial^2 C/\partial z^2$ and therefore it allows higher spatial resolution. Figure S4b shows a representative ac-EFM image at $\omega$ on one of our twisted hBN samples. The potential modulation extends over large regions with regular and irregular triangular domains, similarly as the one observed in dc-EFM in Fig. 1d-f, and only on one side of a monolayer step. Figure S5 shows zoom-in $\omega$ and $2\omega$ images in flat regions around bubbles filled with contamination. The domains are visible in the $\omega$ image, but not in the $2\omega$ image which depends only on the surface dielectric properties. This confirms that the triangular domains reflect a built-in potential that originates at the interface between twisted hBN crystals, not a change in dielectric properties. Figure S6 shows additional images taken in regions around other monolayer steps. Again, they show small triangular domains only on one side of monolayer steps.

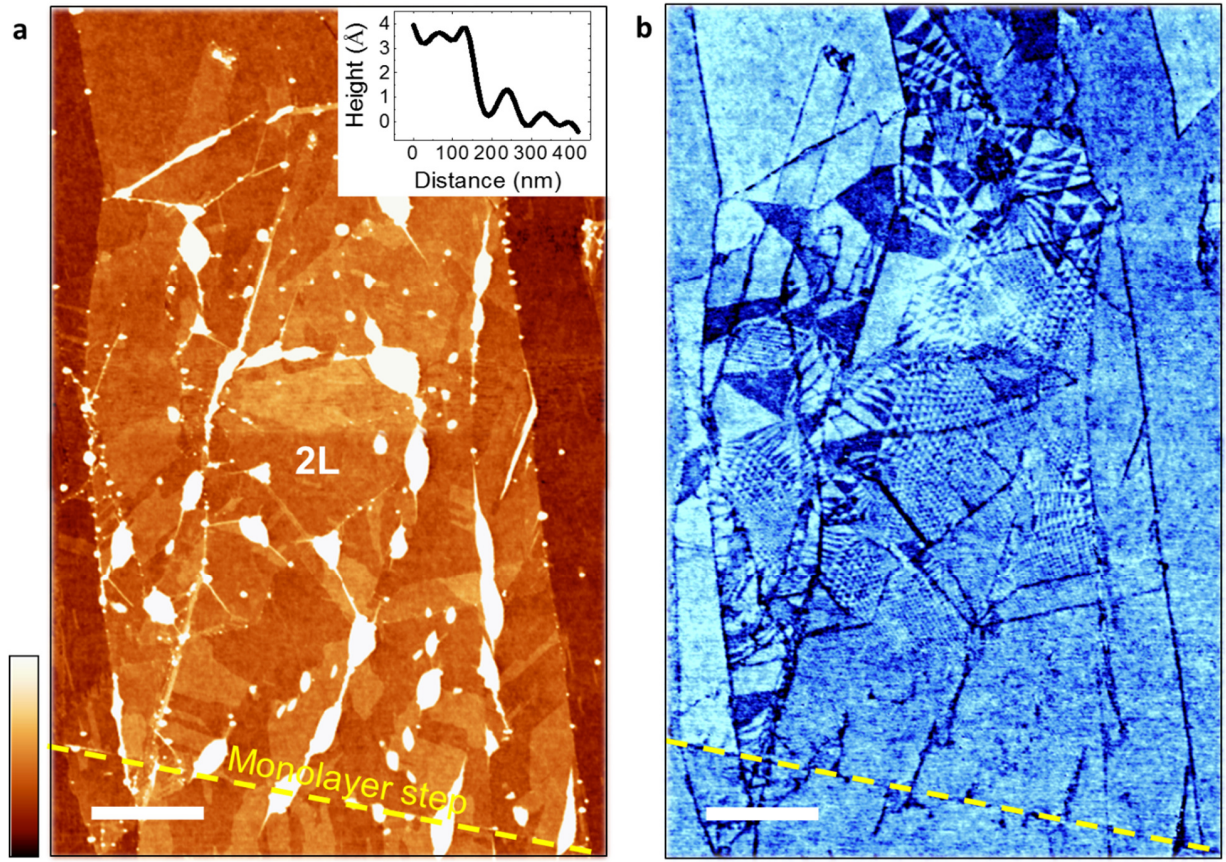

**Figure S4 | ac-EFM image of triangular potential modulation in marginally twisted hBN.** (a) Representative AFM topography of a twisted-hBN sample and **(b)** corresponding ac-EFM image at $\omega$. The colour scale in **(a)** was adjusted to show the details of flat regions instead of bubbles (white regions) that are filled with contamination and several nm high. Large areas with triangular potential modulation are visible in **(b)**, only on the flat regions and only on one side of a monolayer step marked by the yellow dashed lines. The inset in **(a)** shows a profile perpendicular to the monolayer step depicted by the yellow dashed line. Scalebar: **(a,b)** 1.5 μm. Colour bar: 4 nm.

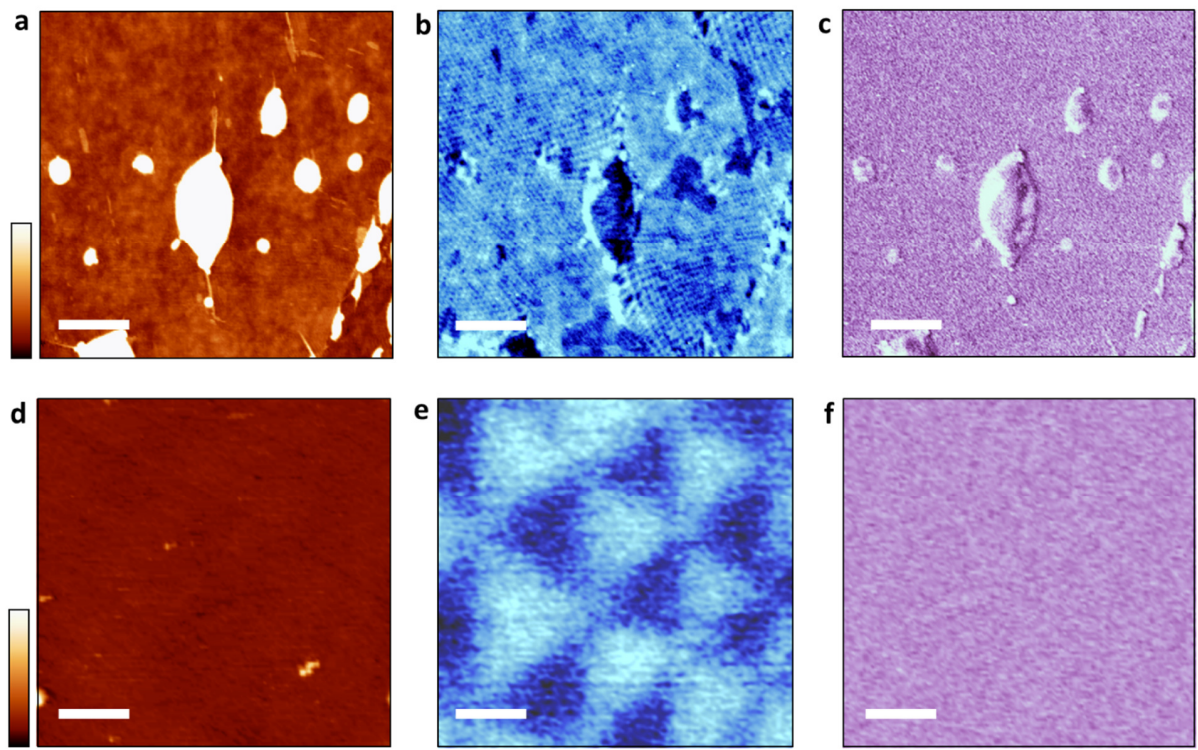

**Figure S5 | ac-EFM ω and 2ω images in marginally twisted hBN. (a,d)** Zoom-in AFM topography images of Fig. S4 and **(b,e)** corresponding ac-EFM images at $\omega$ (surface potential image) and **(c,f)** ac-EFM images at $2\omega$ (dielectric image). A triangular potential modulation is visible in the flat regions only in in the ac-EFM images at $\omega$ in **(b,e)**. No contrast was detected in the dielectric images in **(c,f)**. Scalebar: **(a-c)** 500 nm; **(d-f)** 200 nm. Colour bar: **(a,d)** 2.5 nm.

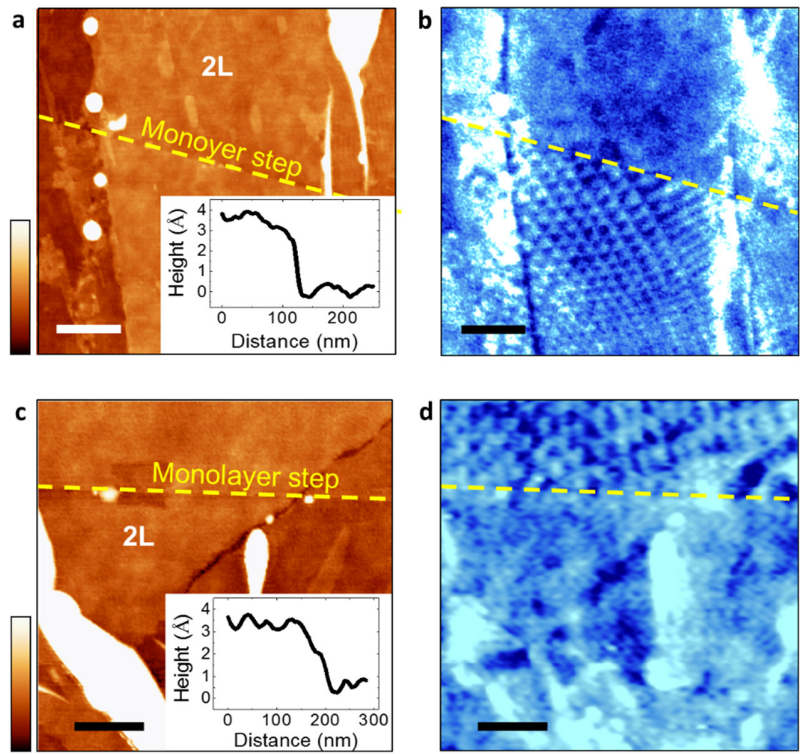

**Figure S6 | Additional EFM images of potential modulation at monolayer terraces. (a,c)** AFM topography images in twisted-hBN in regions around monolayer steps, marked by the yellow dashed lines, and **(b,d)** corresponding ac-EFM images at $\omega$. Insets: the step profiles. A triangular potential modulation is observed only on one side of the steps. Scalebar: **(a,b)** 300 nm, **(c,d)** 200 nm. Colour bar: **(a,c)** 3 nm.

## S3 Experimental quantification of potential modulation and KPFM imaging

We quantified the triangular potential modulation, $\Delta V_s$, in Fig. 3f by measuring the AFM phase shift, $\Delta\varphi$, as a function of the applied dc bias in the centre of two neighbouring domains, as shown in Fig. S7. The observed $\Delta\varphi(V)$ curves (Fig. S7c), which were acquired at the same scan height, are parabolic with same curvature set by the capacitive term $\partial^2 C/\partial z^2$. On the other hand, the maximum of the parabola shifted horizontally with the domain polarity, indicating a change in surface potential[44-46]. We thus quantified $\Delta V_s$ as the difference between the maximum of the two parabolas. The horizontal shift of the parabola also explains the contrast inversion upon changing the sign of the dc bias (Fig. S7a and b). We found $\Delta V_s$ = 240 ± 30 mV on all our samples. The value was robust against variations in the domain shape, orientation and size, from micrometre range down to 30 nm, the smallest domain we could detect with our resolution. We note that for largest domains (1 $\mu$m range) with irregular shape, the domain size in Fig. 3f indicates the smallest side of the domain. The extracted $\Delta V_s$ was also independent of the number of layers in the hBN crystals (within the range of thickness studied here) and of the AFM tip that we used.

It is important to note that the $\Delta\varphi(V)$ curves are sensitive to the height at which they are recorded. This is evident in the curvature of the parabola that changes with the tip-surface distance. The measured potential may also change due to long-range forces from the tip cone and cantilever, which reduce the extracted $\Delta V_s$ if the tip is sufficiently far away from the surface. Such effect is obviously more pronounced with decreasing the size of the domains. To study domains as small as 30 nm and avoid tip-sample convolution effects, all the data in Fig. 3f were taken close to the surface, in the range 8-12 nm from the interface where the dipolar charge distribution is located. The curves were thus taken at a distance smaller than the smallest domain size that we studied. We also employed doped silicon tips (radius < 7 nm, smaller than domain size) instead of conventional metal-coated tips (radius > 25 nm). We then verified that in our experiments the observed $\Delta V_s$ was constant within our accuracy with no significant long-range effects for domain size as small as 30 nm. We note that if larger tips are used or top crystals thicker than a few nm are employed, the observed $\Delta V_s$ is expected to be attenuated for domains smaller than 100 nm.

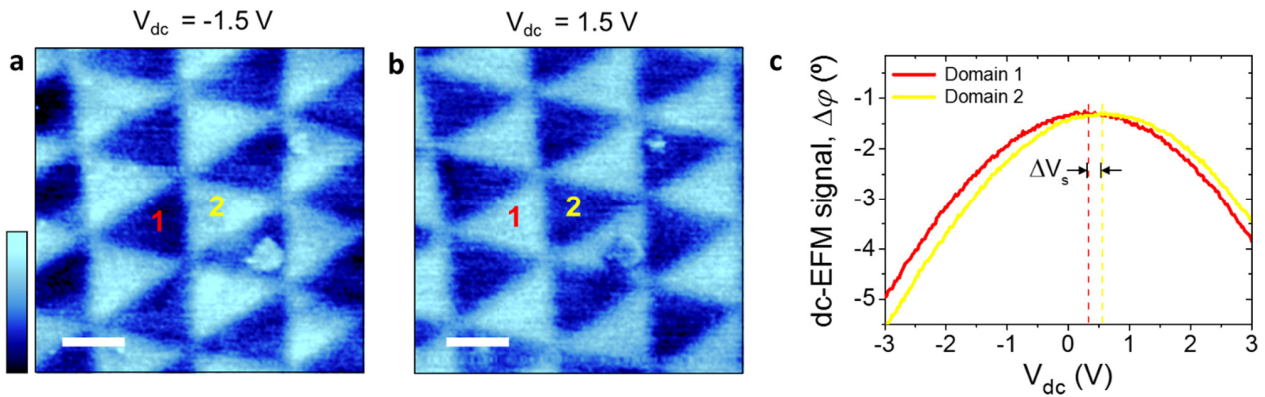

**Figure S7 | Experimental potential variation using dc-EFM. (a,b)** Representative dc-EFM images on marginally twisted hBN with -1,5 V and +1.5 V dc bias applied between the tip and the sample substrate. The triangular contrast reverses upon changing the sign of the dc bias. **(c)** dc-EFM signal, $\Delta\varphi$, as a function of the applied dc bias in the centre of two neighbouring domains in **(a)** and **(b)**, marked as 1 and 2. The two curves show a parabolic behaviour with same curvature, but they are shifted in the horizontal direction, which indicates a variation in the surface potential. The dashed lines are a guide to the eye. Scalebar: **(a,b)** 150 nm. Colour bar: 5°.

To corroborate the observed $\Delta V_s$, we also quantified its value by KPFM imaging[20]. To do that, we used the ac-EFM setup described above (Fig. S3b) with an additional feedback loop and a dc bias between the tip and the silicon substrate. While the tip is scanning, the dc bias is continuously adjusted by the feedback to nullify the amplitude of the ω harmonic, now equal to $F_{el}(\omega) = \partial C/\partial z\, (V_{dc} - V_s) \cdot V_{ac}$. The KPFM image thus yields the surface potential $V_s$ of the sample, mapping its variation across all the domains, not only in their centre. We note that also in KPFM we recorded the phase shift $\Delta\varphi(\omega)$ instead of the force to increase our lateral resolution. Figure S8 shows representative KPFM images of the twisted-hBN sample shown in Fig. 1c, taken at the same scan height as in Fig. 3f (~ 9 nm). We found the same periodic pattern in KFPM images as in the EFM images, with large areas of regular (Fig. S8b) and irregular (Fig. S8a) triangular domains. The KPFM profiles (Fig. S8c and d) show potential variations of 220-270 mV, in agreement with the value extracted from dc-EFM curves.

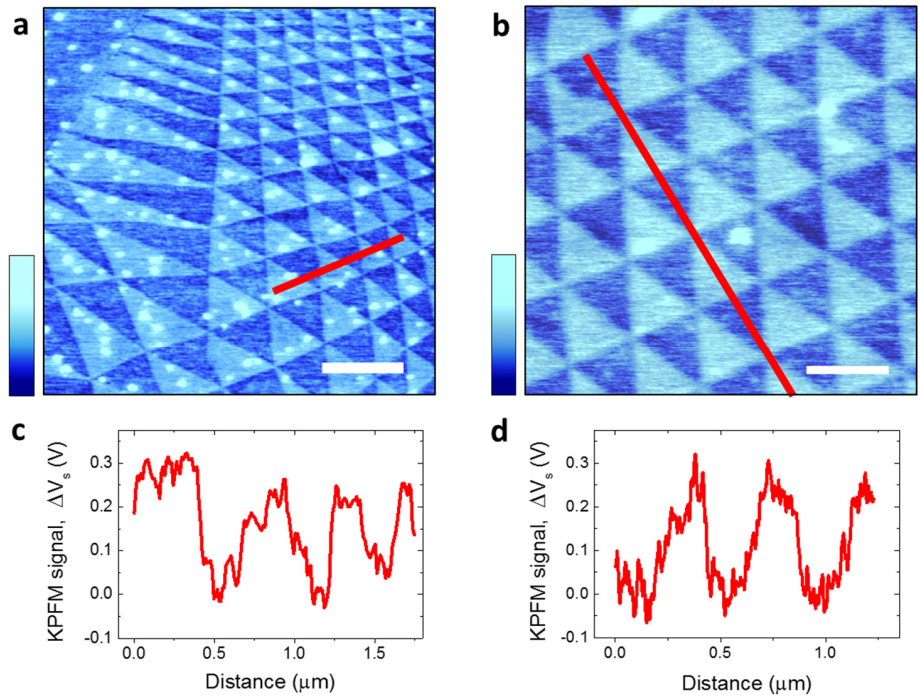

**Figure S8 | Experimental potential variation using KPFM. (a,b)** KPFM images taken on the twisted-hBN sample in Fig. 1d-f in the main text. The same triangular potential pattern is detected as with dc-EFM. **(c,d)** KPFM profiles taken along the red lines in **(a)** and **(b)**, respectively. The surface potential variation between the centres of two neighbouring domains agrees with the value obtained in Fig -3f from $\Delta\varphi(V)$ curves. Acquisition parameters: oscillation amplitude 6 nm; lift height $z_{lift}$ = 3 nm; ac voltage bias of 4 V at 7 kHz. Scalebar: **(a)** 1 μm, **(b)** 250 nm. Colour bar: **(a)** 0.85 V; **(b)** 0.65 V.

## S4 Theoretical calculations

Theoretical results, with an additional analysis of the band structure, will be presented in Ref.[35] Here we shall just present the relevant results from that work.

**Relaxation.** The theoretical calculations were performed in two stages: first, we use LAMMPS[39] to minimize the energy using a classical potential model for relaxation[34], using the 'inter-layer potential'(ILP) from Refs.[40,41] with the Tersoff in-layer potential[42,43]. We minimise the positions for a supercell commensurate with the hBN one, keeping the size of the supercell fixed. Alignments are plotted using an extension of the method in Ref. [34] where we take into account all six alignment options (see Fig. 2 in the main text). This leads to a strain in all these systems that is concentrated along the zone boundaries, and gives rise to a piezoelectric charge, as shown in Fig. S9.

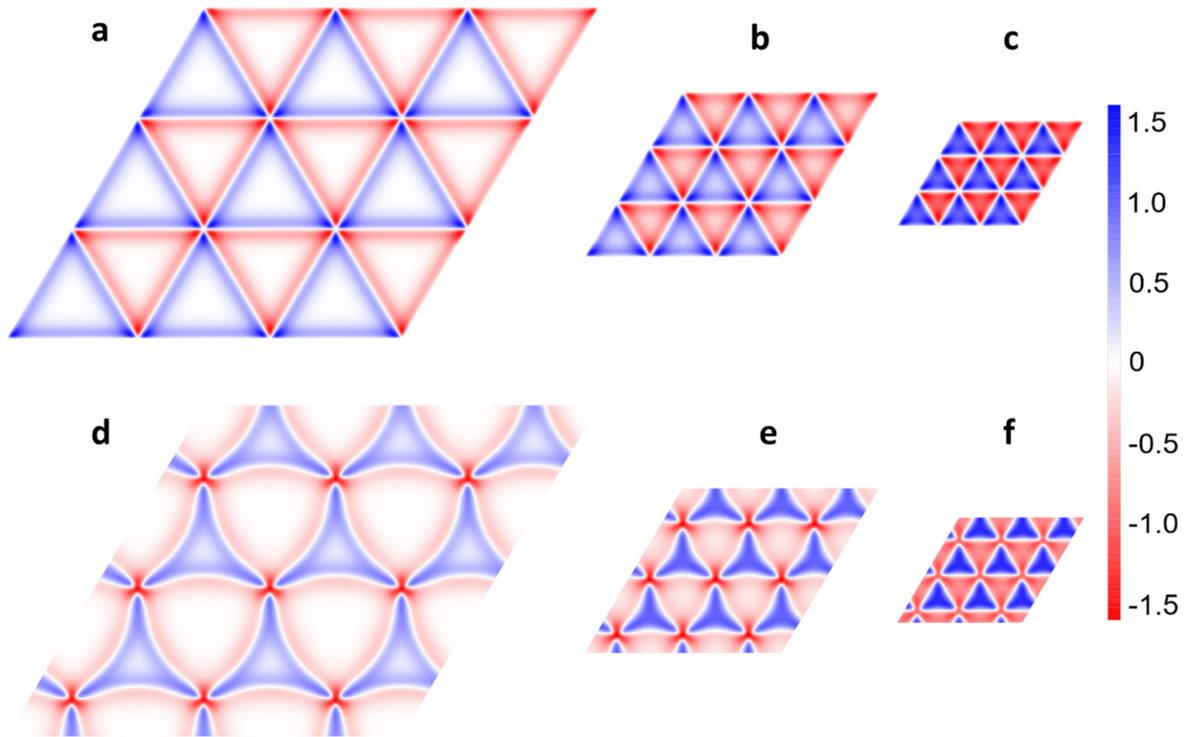

**Figure S9 | Piezo-electric charge in a relaxed hBN bilayer.** The induced piezo-electric charge in a single layer after relaxation of a hBN bilayer: **(a-c)** aligned at angles **(a)** 0.33°, **(b)** 0.67° and **(c)** 1.05°; **(d-f)** anti-aligned at angles **(d)** 0.33°, **(e)** 0.67° and **(f)** 1.05°. The electron density $n$ (scale on the right) is given in units of $10^{12}$ cm$^{-2}$. All images are drawn to the same scale.

**Tight-binding model.** Using the deformed positions, we then perform a tight-binding model. We neglect the modification of the in-layer hoppings due to the small bond stretching, and use a constant in-layer nearest neighbour hopping $t_0 = 2.33$ eV [32]. We use a simple electronic coupling using an exponential Koster-Slater interlayer model,

$$t_{XY}(r) = t_{XY} \exp(-\alpha\,(r-d)),$$

where $X$ and $Y$ label the atomic species, $d = 0.333$ nm is the interlayer distance, and the inverse range $\alpha = 44$ nm$^{-1}$.

We then diagonalize the resulting tight-binding model, either for energies near the gap (which allows us to use sparse matrix techniques, and thus study much larger moirés) or by finding all states, which is required to describe the charge density. This is calculated by summing over all occupied states, which limits the smallest angle we can perform calculations for to about 1°. Further details of these calculations, as well as further results on the electronic structure of hBN can be found in Ref.[35].

We note that in contrast to the experimental results, all our theoretical results are for a bilayer system. However, we have shown[47] that relaxation in multilayer systems still shows a sizeable reconstruction at the interface. Nevertheless, we expect that the results we get for the piezoelectric charge may be a substantial overestimate of their real magnitude.

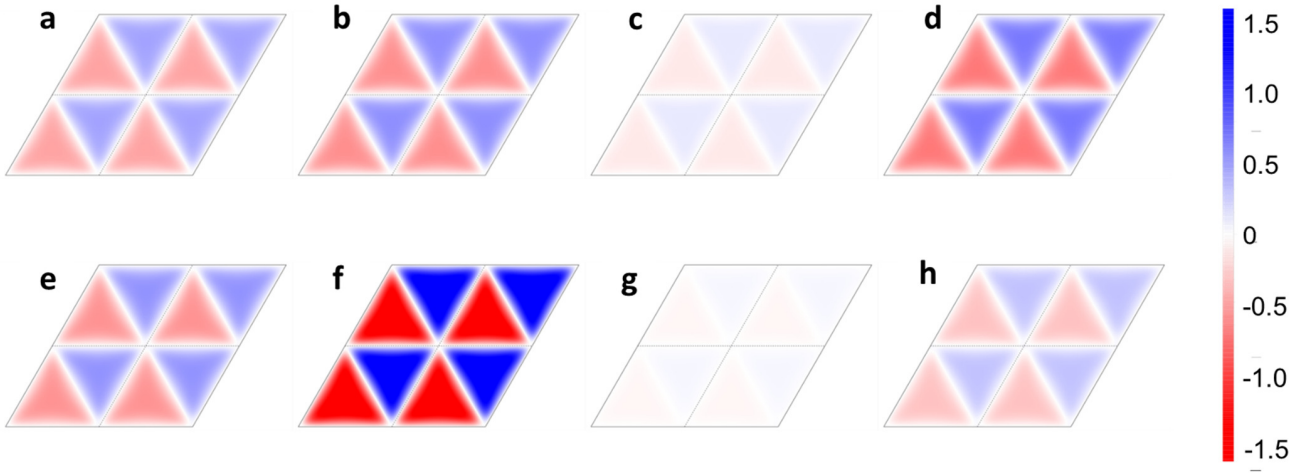

**Figure S10 | Charge density in the top-layer of twisted hBN for parallel alignment.** Twist induced charge density at neutrality in the top layer for $\theta = 1.05°$ for a relaxed layer for change to the basic parameter choice $t_{BB} = 0.7$, $t_{NB} = 0.3$, $t_{NN} = 0.15$, $\Delta = 8$ eV, $\alpha = 44$ nm$^{-1}$ **(a)** basic parameters; **(b)** $t_{BB} = 0.5$ eV; **(c)** $\alpha = 22$ nm$^{-1}$; **(d)** $\alpha = 66$ nm$^{-1}$; **(e)** $\Delta = 6$ eV; **(f)** $t_{NB} = 0.5$ eV; **(g)** $t_{NB} = 0.15$ eV; **(h)** $t_{BB} = 1.0$ eV. The units are the same as in Fig. S9 ($10^{12}$ cm$^{-2}$).